\documentstyle[floats,aps]{revtex}
\input{psfig}
\begin{document}
\def \beq{\begin{equation}}
\def \eeq{\end{equation}}
\def \beqarr{\begin{eqnarray}}
\def \eeqarr{\end{eqnarray}}
\def \be{\begin{equation}}
\def \ee{\end{equation}}
\def \bea{\begin{eqnarray}}
\def \eea{\end{eqnarray}}
\def \ta{{\tilde\alpha}}
\def \tg{{\tilde g}}     
\twocolumn[\hsize\textwidth\columnwidth\hsize\csname @twocolumnfalse\endcsname
\title{Edge States on a Quantum Hall Liquid--Solid Interface}
\author{Milica Milovanovic$^{1,2}$ and Efrat Shimshoni$^2$}
\address{$^1$ Department of Physics, The Technion, Haifa 32000, Israel.}
\address{$^2$ Department of Mathematics-Physics, Oranim--Haifa University,
Tivon 36006, Israel.}
\date{\today}
\maketitle
\begin{abstract}
We study the edge--states excitations of a droplet of quantum Hall liquid 
embedded in an electron (Wigner) solid. The presence of strong correlations
between the liquid and solid sectors in the ground state is shown to be
reflected in the density of states $D(E)$, 
associated with the excitations of the liquid--solid interface.
We find that the prominent effect of these correlations is a
suppression  of $D(E)$ with respect to its value ($D_0(E)$) in the absence of
the Wigner solid environment:  $D(E) \sim e^{-\alpha |E|}D_0(E)$. 
The coefficient
$\alpha$ (which is shown to vanish for a perfectly regular 
distribution of electron sites in the solid), is evaluated for two different
realizations of an irregular distribution. We conclude that probing
this effect (e.g. in a tunneling experiment), can provide
evidence for correlated liquid--solid mixture states in
quantum dots, or disordered samples, in very strong magnetic fields.
\end{abstract}
\pacs{73.40.Hm, 72.30.+q, 75.40.Gb}
\vskip2pc]
\narrowtext
\section{Introduction and Principal Results}
\label{sec:intro}
The two--dimensional electron gas (2DEG) in strong perpendicular
magnetic fields can form a variety of exotic quantum phases. In
particular, in clean systems at moderately low filling fractions (close to 
$\nu=1/5$), the correlations which favor a fractional quantum Hall liquid 
(QHL) state compete with the crystalline order of a Wigner solid (WS)
\cite{ws,Herb}. This competition can induce transitions between the QHL state
and the insulator, as has been observed experimentally \cite{exp1/5}. 
In the presence of slowly varrying disorder or a confining potential, the 
electronic ground state may develope a fractured order -- namely,
form a binary liquid--solid mixture. A liquid--solid separation
possibly occurs also at higher filling fractions (in which the electrons
in the solid regions should form a glassy state controlled by short--range 
disorder). An experimental evidence for this scenario is provided by 
photoluminescence data \cite{Kuk}. In addition, a set of puzzling
transport data \cite{nuc,Shahar} can be explained most naturally under
the assumption of a macroscopic inhomogeneity \cite{macinhom}.

The interplay of QHL correlations and crystallization in the low $\nu$
regime has been clearly demonstrated by Zheng and Fertig \cite{Herb}.
Using a variational calculation, they have shown that a Wigner lattice
with an interstitial electron introduced via a Laughlin--like Jastrow factor,
can be lower in energy compared to the perfect WS with the same total number
of electrons. This implies that in a certain range of $\nu$'s, the
crystal is unstable to a specific type of density fluctuations - pre--formed
QHL droplets. In the presence of density fluctuations induced by a
slowly varrying external potential, it is reasonable to expect a
nucleation of such interstitials in the higher density regions. It is therefore
suggestive that the ground state slightly below $\nu=1/5$ separates
into QHL and WS sectors, which are correlated by a Jastrow factor to
minimize the energy of electrons close to the liquid--solid interface (see
Eq. (\ref{psiLS})).

In the present paper we investigate the physical implications of 
liquid--solid mixture states, as reflected by the corresponding
low lying excitations. Similarly to a finite
droplet of a primary QHL (of $\nu=1/m$ with $m$ an odd integer), the
gapless excitations are chiral edge states \cite{Wen}, i.e.,
deformations of the boundary of the incompressible droplet, which travel
in a definite direction along the boundary. However, in case the liquid droplet
is embedded in an electron solid (rather than a vacuum), the nature of
these excitations of the liquid--solid interface is affected by the
correlations between the two sectors. In particular, high amplitude
deformations of the interface are generally suppressed, since the
liquid electrons are constrained by their tendency to avoid the proximity
to localized sites of the WS as much as possible. 
This can lead to a {\it decay} of the density of
states with increasing deviation from the Fermi level, as long as higher energy
excitations (which involve, e.g., a reorganization of the WS electrons)
are not yet activated. 

To facilitate the derivation of this peculiar effect, we consider a simple
geometry of a large, circular quantum dot, in which the electrons are assumed
to form a disc of QHL surrounded by a WS (see Fig. 1). We evaluate the electron
propagator and consequently the density of states for tunneling
into the liquid--solid interface, $D(E)$. In the thermodynamic limit, we
find
\beq
D(E)\sim e^{-\alpha |E|}D_0(E)\; ,
\label{dos}
\eeq
where $D_0(E)$ corresponds to the ordinary edge--state (on an interface 
between QHL and a vacuum), and the coefficient $\alpha$ depends crucially on 
the distribution of localized sites in the WS sector. In particular, when 
these sites form a structure with a perfect crystaline order arround the disc 
(and the lattice constant is commensurate with the circumference),  
$\alpha=0$: in that case, the WS electrons merely deform the effective 
boundary of the liquid into a regular shape, as depicted in Fig. 1(a). 
In contrast, an irregular distribution of sites induces frustration, and thus a
suppression of $D(E)$.

The most direct way to probe this effect is via tunneling into the 2DEG, e.g.
using the technique developed by Ashoori \cite{Ashoori}: the tunneling
conductance is given by $G(V)\sim D(eV)$ (where $V$ is the voltage across the 
tunnel barrier). The suppression of $G(V)$ at low $V$ may lead to a 
non--monotonous behavior -- at higher voltage bias, higher energy excitations
take over and induce an increase of $G(V)$. The effect is expected to become 
more pronounced with increasing inhomogeneity of the external potential. In
particular, in a quantum dot where both the number of electrons and the 
confining potential can be controlled, the suppression of $G(V)$ is
expected to exhibit oscillations: the tunneling rate should be maximized when 
the control parameter enables a nearly regular configuration of sites in the
WS sector.

In the following sections, we detail the derivation of the electron propagator 
along the liquid--solid interface (Sec. \ref{sec:ep}), and the implied 
behavior of the density of states, Eq. (\ref{dos}) (Sec. \ref{sec:dos}). 
In the latter, we
consider two different realizations of the irregularity in site configuration:
(a) a regular distortion of the circular symmetry, and (b) a symmetric
random distribution. The corresponding expressions for the 
suppression--time $\alpha$ are given by  Eqs. (\ref{alphih}) (case (a)) and 
(\ref{alphrp}),(\ref{alphrr}) (case (b)).

\section{Derivation of the Electron Propagator}
\label{sec:ep}
As we explained in the introduction we expect that the ground state wave
function that describes the QH droplet surrounded by WS electrons is 
essentially of the following form:

\beq
\Psi_{LS}= {\cal A} \{\Psi_{L}(z_{1},\ldots ,z_{N}) 
\Psi_{S}(w_{1},\ldots,w_{M})\prod_{i,j} (z_{i} - w_{j})^{m}\}
\label{psiLS}
\eeq
where $\Psi_{L}$ describes the liquid part, $\Psi_{S}$ the solid part, and the
last expression describes the Jastrow correlations of these two phases. 
$ {\cal A} $ denotes antisymmetrization over all electron coordinates. 
$ m $ corresponds to the filling factor $ \nu = 1/m $ of the QH liquid part:

\beq
\Psi_{L} (z_{1},\ldots, z_{N}) = \prod_{i < j} (z_{i} - z_{j})^m
 \exp\{-\frac{1}{4} \sum |z_{i}|^{2}\}
\label{psidisk}
\eeq
The wave function
\beq
\Psi_{S} (w_{1}, \ldots, w_{M}) = \exp \{\frac{1}{2} \sum \overline{w}_{i}
 R_{i} - \frac{1}{4} \sum |w_{i}|^{2}\}
\label{psiws}
\eeq
describes WS electrons localized on the positions $ R_{i}, i=1,\ldots,M $. 
(It is a multiple of lowest--Landau--level delta functions.)

Our main assumption in the derivation of the electron propagator is that the 
edge of the QH droplet, coupled to WS, essentially behaves  as a slightly 
modified Luttinger liquid. 
That assumption allows us to use a construction of one
electron state on the edge similar to the one of the Luttinger liquid. At the 
end of the derivation we will be able to specify constraints on the 
configuration of the WS sites, such that the assumption is valid.

Because of the above assumption, it would be instructive first to briefly 
recapitulate the derivation of the (equal-time) propagator when the QH disc is
surrounded by vacuum \cite{Wen}.

The derivation begins by considering $m$ - Laughlin quasihole constructions, 
i.e.
\beq
\prod_{i=1}^{N} (z_{i} - \xi)^{m} \Psi_{L} = \Psi_{o}(\xi)
\label{quasihole}
\eeq
where $\xi$ lies outside of the system; $ |\xi| > R $, and $R = 2 m N$ is the 
radius of the QH droplet. $ \Psi_{L} $ is the Laughlin wave function (Eq.
(\ref{psidisk})). The first step towards the electron correlator is calculating
the following scalar product of the state (\ref{quasihole})
\beq
{\cal N} (\xi^{*}, \xi) = 
\frac{<\Psi_{o}(\xi)|\Psi_{o}(\xi)>}{<\Psi_{L}|\Psi_{L}>}\xi^{-2mN}\; .
\label{scalprod}
\eeq
The leading contribution in the $ R/|\xi|$ expansion can be obtained by the 
plasma analogy \cite{Wen} or simply by considering the decomposition of 
coordinates into the center of mass, $ Z_{cm} = \frac{1}{N} 
\sum_{i=1}^{N} z_{i}$, and relative ones. When the numerator is approximated 
by 
\begin{eqnarray}
  m \sum_{k} \ln |z_{k} - \xi|^{2} \approx & \nonumber \\
& N m 2 \ln |\xi| - m N \frac{Z_{cm}}{\xi} - m N \frac{Z_{cm}^{*}}{\xi},
\label{expansion}
\end{eqnarray}
the integration over the the center of mass coordinate is decoupled from the 
other integrations which do not depend on $\xi$. It
yields the leading dependence on $ R/|\xi|$:
\beq
{\cal N}_{o} (\xi^{*}, \xi) \approx (1 + m \frac{R^{2}}{|\xi|^{2}})\; .
\label{singularity}
\eeq
In an average, macroscopic picture, we expect singular behavior as 
$ |\xi| \rightarrow R$, with singularity at $ |\xi| = R $ (for the equal-time,
equal-space correlator) \cite{Wen}. Then Eq. (\ref{singularity}) can 
be rewritten as
\beq
{\cal N}_{o} (\xi^{*}, \xi) = (1 -  \frac{R^{2}}{|\xi|^{2}})^{-m}
\label{llform}
\eeq
and we will assume that it is valid also for $ |\xi| \sim R \; \; (|\xi|> 
R) $. By doing this we neglect any finite size corrections (due to finite 
$ R $) which might be present in $ {\cal N}_{o} $ as $ |\xi| \rightarrow R $. 
  To get the electron propagator we analytically continue the function 
$ {\cal N}_{o} (\xi^{*}, \xi) $ of the variable $ \xi $ to $ {\cal N}_{o} 
(\tilde{\xi}^{*},\xi) $ which depends on $ \tilde{\xi} $ and $ \xi $. 
That allows us to take $\xi$ and $ \tilde{\xi}$ to the edge of the system - 
 $ \xi = R \exp\{i 2 \pi \frac{x}{L}\} $ and $ \tilde{\xi} = R $, without 
encountering the singularity for $ x=0 $ (which determines the behavior of 
the function in its neighborhood). By taking $\xi$ and $ \tilde{\xi} $ to the 
edge, we in fact describe a particle-hole excitation on the edge that goes 
into the electron propagator, and find
that for $ x \ll L $ the propagator behaves as
\begin{eqnarray}
 &  & G_{e}^{o}(x) = {\cal N}_{o} (\tilde{\xi}^{*},\xi) \tilde{\xi}^{* (N - 1)
 m} \xi^{(N - 1)m} \sim \nonumber \\
 &  & \; \; \; \; \; \frac{1}{(\frac{x}{L})^{m}} \exp\{i m (N - \frac{1}{2}) 
\frac{2 \pi}{L} x\}
\label{llpro}
\end{eqnarray}
$ m (N - \frac{1}{2}) \frac{2 \pi}{L} $ in the exponential is the value of the
generalized $(m \neq 1)$ ``Fermi momentum''.

Following a similar strategy, we address the case of the droplet surrounded 
by WS, described by the wave function (\ref{psiLS}). 
We consider the following state
\beq
\Psi(\xi) = \prod_{i=1}^{N} (z_{i} - \xi)^{m} \Psi_{LS}
\eeq
and the corresponding scalar product:
\beq
{\cal N}(\xi^{*},\xi) = \frac{<\Psi(\xi)|\Psi(\xi)>}{<\Psi_{LS}|\Psi_{LS}>}
\xi^{-2mN}
\eeq
where
\begin{eqnarray}
& &<\Psi(\xi)|\Psi(\xi)> =  \nonumber \\
& &
\int \prod_{i=1}^{N} d^{2}z_{i} \int \prod_{j=1}^{M} d^2w_{j} 
\exp\{\sum_{i<j} 2m \ln |z_{i} - z_{j}|\nonumber \\
& & 
+ \sum_{i=1}^{N} \sum_{j=1}^{M} 2m \ln |z_{i} - w_{j}| - 
\frac{1}{2} \sum_{i=1}^{N} |z_{i}|^{2}  \nonumber \\
& &- \frac{1}{2} \sum_{j=1}^{M} |w_{j} - R_{j}|^{2} + 2m \sum_{i=1}^{N} 
\ln |z_{i} - \xi|\}\; .
\label{plasmaint}
\end{eqnarray}
In the above formulas the antisymmetrization was neglected, 
which is possible due to the localized nature of the WS electrons. 
Again, if for $ |\xi| \gg R $ and $ |w_{j}| \gg R, \; j = 1,\ldots,M $, 
the last sum in (\ref{plasmaint}) is approximated as (\ref{expansion}), and
\begin{eqnarray}
& &\sum_{i=1}^{N} \sum_{j=1}^{M} 2m \ln |z_{i} - w_{j}| \approx \nonumber \\
& &\sum_{j=1}^{M} 2mN \ln |w_{j}| - m N Z_{cm} \sum_{j=1}^{M} \frac{1}{w_{j}} 
- m N Z_{cm}^{*} \sum_{j}^{M} \frac{1}{w_{j}^{*}},
\end{eqnarray}
the integration over $Z_{cm}$ yields the functional dependence of the 
integrals over $z_i$'s in (\ref{plasmaint}) on $\xi$ to leading order in 
$ R | \sum_{j=1}^{M} \frac{1}{w_{j}} + \frac{1}{\xi}|$. 
It is of the following form:
\beq
(1 + m R^{2} |\sum_{j=1}^{M} \frac{1}{w_{j}} + \frac{1}{\xi}|^{2}).
\label{ftts}
\eeq
Now we will assume that the sites $ R_{j}, \; j=1,\ldots, M $ are such that 
the integration over $w_j$'s is dominated by contributions for which
\beq
R^{2 n} | \sum_{j=1}^{M} \frac{1}{w_{j}^{n}} + 
\frac{1}{\xi^{n}}|^{2} < 1, \; \; n = 1,\ldots, \infty \; .
\eeq
Then, the result of the $z$-integration can be expressed as an expansion in 
variables symmetric in $\xi$ and $w_j$'s, defined as
\beq
x_{n} = \sum_{j=1}^{M} \frac{1}{w_j^{n}} + \frac{1}{\xi^{n}}, \; n = 1, 
\ldots, \infty\; .
\label{sympol}
\eeq
The first two terms of the expansion are given by the expression (\ref{ftts}).
As explained in Appendix A, the final $w$-integration amounts to replacing 
$w_{j}$ with $ \tilde{R}_{j} $, defined in the Appendix, and variables 
(\ref{sympol}) in the expansion yield
\beq
X_{n}\{R_{j},j=1,\ldots,M\} = \sum_{j=1}^{M} \frac{1}{{\tilde{R}_{j}}^{n}} + 
\frac{1}{\xi^{n}}\; .
\label{Sympol}
\eeq
At this point, we can see that in the case where the positions of WS 
electrons satisfy
\beq
\sum_{j=1}^{M} \frac{1}{\tilde{R}_{j}^{n}} = 
\sum_{j=1}^{M} \frac{1}{\tilde{R}^{* n}} = 0,
\; \; n = 1, \ldots, \infty
\label{conditions}
\eeq
the problem reduces to the one of the droplet surrounded by vacuum, 
and the expansion should sum up to the Luttinger-liquid form (\ref{llform}). 
These conditions can be satisfied, e.g., when 
\beq
\tilde{R}_{j} = \tilde{R} \exp\{i (\theta_{j} + \theta_{o})\}
\label{pchain}
\eeq
where $ \theta_{j} = j \theta $, $ \theta = \frac{2 \pi}{M} $, and 
$ \tilde{R} $ is a constant radius, i.e., when a commensurate chain of WS 
electrons surrounds the droplet (see Fig. 1(a)). 
Note that our assumption of a small 
correction to  the Luttinger-liquid behavior is justified, provided the
configuration of WS sites is a small perturbation of one which satisfies
Eq. (\ref{conditions}).

To get the electron propagator, we first exponentiate the expression of the 
leading order behavior for
$ |\xi| \gg R , \; |R_{j}| \gg R, \; j = 1, \ldots, M $
\beq
{\cal N} (\xi^{*},\xi) \approx (1 - R^{2} |\sum_{j=1}^{M} 
\frac{1}{\tilde{R}_{j}}+ \frac{1}{\xi}|^{2})^{-m}.
\label{propagator}
\eeq
Here we assume that even for $R_{j}$'s close to the droplet, 
$ \sum \frac{1}{\tilde{R}_{j}} $ is small with respect to
$ \frac{1}{|\xi|} \sim \frac{1}{R} $, (which implies that the pole structure 
of the correlator is similar to the one of the Luttinger liquid with the pole 
slightly shifted from $|\xi| = R $). Hence, we can also regard Eq.
(\ref{propagator}) valid for general $ R_{j}, \; j = 1, \ldots, M $, and 
$ \xi $ (with $ |R_{j}|, |\xi|> R $). ${\cal N} (\xi^{*},\xi)$ is then
analytically continued to ${\cal N} (\tilde{\xi}^{*},\xi)$;
taking $ \xi = R \exp\{i 2 \pi \frac{x}{L}\} $ and $ \tilde{\xi} = R $ 
gives for the electron equal-time propagator
\begin{eqnarray}
&G_{e}(x) \sim \exp\{i m (N - \frac{1}{2})2 \pi \frac{x}{L}\} \times 
\nonumber \\
& [\exp\{i \pi \frac{x}{L}\} - \exp\{-i \pi \frac{x}{L}\} \nonumber \\
& - \exp\{i \pi \frac{x}{L}\} \Sigma - 
\exp\{- i \pi \frac{x}{L}\} \Sigma^{*}\nonumber \\ 
& - \exp\{i \pi \frac{x}{L}\} | \Sigma |^2]^{-m}\\
& {\rm where}\quad \Sigma\equiv\sum_{j=1}^{M} \frac{R}{\tilde{R}_{j}}
\nonumber 
\end{eqnarray}
In the limit $ \frac{x}{L} \ll 1 $ that corresponds to the short distance 
behavior, the electron propagator can be expressed as
\beq
G_{e}(x) \sim \exp\{i m (N - \frac{1}{2}) 2 \pi \frac{x}{L}\} \times
\frac{1}{(i x + C_{1} x - C_{2})^{m}}
\label{Gex}
\eeq
where
\beq
C_{1}= \frac{2{\cal I}\{\Sigma\}}{2 - |\Sigma |^2}
\eeq
and
\beq
C_{2} = \frac{L}{\pi} \frac{(2{\cal R}\{\Sigma\}+ |\Sigma |^2)}
{2 -  |\Sigma |^2}
\label{c2def}
\eeq
(${\cal R}\{\Sigma\}$ and ${\cal I}\{\Sigma\}$ are the real and imaginary parts
of $\Sigma$, respectively). Note that $|\Sigma |$ serves as a small
parameter that determines the deviation from a standard Luttinger liquid
behavior (Eq.  (\ref{llpro})).

\section{The Tunneling Density of States}
\label{sec:dos}
In the previous section we calculated the equal time propagator. In order to 
get the tunneling density of states, we need the  time--dependent propagator 
$G_e(x=0,t)$. In general, this requires the 
knowledge of the energies of excited states of the system. For WS 
configurations that satisfy the conditions  (\ref{conditions}) (and hence have
the Luttinger liquid correlations (\ref{llpro})), 
we expect that the ground state is
one of the edge states of a QH droplet surrounded by vacuum;
the excited states (which may also be interpreted as edge states of that
droplet) have energies linear in momentum, measured from the new ground state. 
The linear dispersion is expected on general grounds, as the
first order expansion in small momenta, and not precluded by any
(symmetry etc.) argument \cite{fn1}. Then, to get $G_e(x=0,t)$ 
we should merely substitute the coordinate $x$ with $ v t$
(where $v $ is the velocity of the drift motion of electrons on the 
boundary \cite{Wen}). The sign of time should be specified, as we will 
explain and elaborate below. Assuming that the linear  dispersion is valid 
also in the case of a small deviation from Eq. (\ref{conditions})
(i.e., for $|\Sigma |\ll 1$), 
we may apply the same substitution in (\ref{Gex}) to get $G_e(x=0,t)$.

We recall the definition of the (equal-space) fermionic Green's function, 
in the field-theoretic notation:
\beq
G_{e}(x=0,t > 0) = - i <0| \Psi(0,t) \Psi^{\dagger}(0,0)|0>,
\label{postimes}
\eeq
and
\beq
G_{e}(x=0,t < 0) =   i <0| \Psi^{\dagger}(0,0) \Psi(0,t)|0>
\label{negtimes}
\eeq
where $\Psi^{\dagger}$ and $\Psi$ are the electron and hole creation operators,
respectively. If we assume that, in our system, equation
\beq
<0|\Psi(0,t) \Psi^{\dagger}(0,0)|0> = <0| \Psi^{\dagger}(0,t) \Psi(0,0)|0>
\eeq
holds, as it does in the case of the standard Luttinger liquid, 
going from formula (\ref{negtimes}) to formula (\ref{postimes}) 
involves only time
translation $(T \rightarrow T-t)$ and time inversion $(T \rightarrow - T)$ (and
the overall change of the sign). In our case, the particle coordinates
$\xi = R \exp\{i x \frac{2 \pi}{L}\}$ and $\tilde{\xi} = R$ become under the 
substitution ($x \rightarrow v t$) 
$ \xi(t) = R \exp\{i vt \frac{2 \pi}{L}\} $ and
$ \tilde{\xi}(0) = R $. $\xi(t)$ and $\tilde{\xi}(0)$ denote the hole and 
electron coordinate respectively.
If we reconsider the correlator ${\cal N}(\tilde{\xi}^{*},\xi)$ with the
substitutions, we will get for the hole propagator 
(\ref{negtimes}) in the short time limit:
\beq
G_{e}(0,t < 0) \propto \frac{1}{(i v t + C_{1} v t - C_{2})^{m}}
\label{neggreen}
\eeq
On the other hand, to get the electron propagator for $ t > 0 $ 
(\ref{postimes}) we should perform the time translation and inversion. 
This amounts to an exchange of $\xi(t)$ and $\tilde{\xi}(0)$, which leads to 
the following short time behavior:
\beq
G_{e}(0,t > 0) \propto \frac{1}{(- i v t + C_{1} v t - C_{2})^{m}}
\label{posgreen}
\eeq
If we consider $t$ as a complex variable and assume $ |\Sigma| \ll 1 $,
we may approximate the positions of poles in
(\ref{neggreen}) and (\ref{posgreen}) as $ t_{1} \approx - i \frac{C_{2}}{v}
$, and $ t_{2} \approx i \frac{C_{2}}{v}$, respectively.
Then, the tunneling density of states $D(E)$ is given by 
\begin{eqnarray}
& D(E) \sim & {\cal{R}} \{\int_{- \infty}^{0}  dt \; \exp\{i E t\} 
\frac{1}{(it - C_2/v)^{m}} + \nonumber \\
& &                      \int_{0}^{\infty}  dt \; \exp\{i E t\}
\frac{1}{(it + C_2/v)^{m}} \}
\label{dosint}
\end{eqnarray}
where $E$ is the energy measured from the Fermi energy, $\hbar=1$. 
As a final result we get 
\beq
D(E) \propto \exp\{- \alpha |E|\} D_{o}(E) \; ,\quad
\alpha\equiv\frac{|C_{2}|}{v}\; ,
\label{dosalph}
\eeq
where $ D_{o}(E)$ is the Luttinger liquid tunneling density of states.
Therefore,  to lowest order in  $|\Sigma |$,
 the dominant modification to the standard
 Luttinger behavior is the exponential suppression, at a time scale $\alpha$ of
order $\sim(L/v)|\Sigma |$.  
It should be stressed that  $ \alpha $, and hence $D(E)$, 
is a local quantity (adiabatically varrying around the disc)
that describes the tunneling density of 
states of an electron at  distance $ \sim R $ from the origin, and at
angle $\pi \frac{x}{L} \ll 1$ to the reference point.
 
Below we calculate $\alpha$ for the two different types of 
imperfect site configurations depicted in Fig. 1(b),(c). 
\bigskip

\centerline{{\bf A. Inhomogeneous Configuration of Sites}}
We model the inhomogeneous configuration (Fig. 1(b)) by considering 
$ M = 2n + 2  $ WS electrons, at a distance $\tilde{R}$ from the origin, 
where two of them are exactly on the opposite sides of the droplet, 
i.e. the sum of their phases is 
$  \exp\{i \pi \} + 1 = 0 $. The rest $2n$ electrons are positioned 
at the angles $\theta_j$ and $-\theta_j$ ($1\leq j\leq n$), where
\beq
\theta_j= \theta j + \epsilon \cos\{\delta (j-1)\}\; , 
\eeq
$ \theta = \frac{\pi}{n + 1} $ and $ \delta = \frac{\pi}{2 n} $. 
$\epsilon > 0 $ represents a small deviation from the perfect chain 
distribution, which is modulated, as we move from the reference point 
at angle $\gamma=0$. We then get $\Sigma=\Sigma_{0}$, where
\begin{eqnarray}
& \Sigma_{0} \equiv & \frac{R}{\tilde{R}}\sum_{j=1}^{n} 
\exp\{i \theta j + i \epsilon \cos\{\delta (j-1)\}\} + \nonumber \\
& & \frac{R}{\tilde{R}}\sum_{j=1}^{n} \exp\{- i \theta j 
- i \epsilon  \cos\{\delta (j-1)\}\}\; .
\end{eqnarray}

For $\epsilon$ small $ \Sigma_{0} $ can be approximated as
\begin{eqnarray}
& \Sigma_{0} \sim  & i \epsilon \frac{R}{\tilde{R}} \sum_{j=1}^{n} 
\exp\{i \theta j\} \cos\{\delta (j-1)\}  \nonumber \\
&   &- i \epsilon \frac{R}{\tilde{R}} \sum_{j=1}^{n} \exp\{- i \theta j\}
\cos\{\delta (j-1)\}
\end{eqnarray}
and for $n\gg 1$ this yields
\beq
\Sigma_{0} = - \frac{8 n \epsilon}{3\pi} \frac{R}{\tilde{R}}\; . 
\label{sigzero}
\eeq

If the reference point is at an arbitrary angle $ \gamma \neq 0 $, 
the sum becomes
\beq
\Sigma = \Sigma_{0} \exp\{- i \gamma\}\; .
\eeq
Inserting in  Eqs. (\ref{c2def}) and (\ref{dosalph}), we then get
\beq
\alpha = \frac{L}{\pi v}\left| \frac{2 \Sigma_{0} \cos\{\gamma\} + 
|\Sigma_{0}|^{2}}{2 - |\Sigma_{0}|^{2}}\right|\; .
\label{alphih}
\eeq
Therefore, at $ \gamma = \frac{\pi}{2} $, $\alpha$ becomes $ \alpha \propto
\epsilon^{2} $, at $\gamma = 0 $ it is 
$\alpha \propto (\epsilon - {\rm Const} \;
\epsilon^{2})$, $ {\rm Const} > 0 $, and, at $\gamma = \pi$, it is
$\alpha \propto (\epsilon + {\rm Const} \; \epsilon^{2})$.
This means that around $\gamma = 0$ and $\gamma = \pi$ deviations from the 
perfect chain case are stronger, and the suppression of the density of states
is stronger in the denser region $(\gamma = \pi)$, but only to second order
in $\epsilon$.
\bigskip

\centerline{{\bf B. Random Distribution of Sites}}
We consider a configuration of sites obtained by a random distortion of the 
perfect chain Eq. (\ref{pchain}) (see Fig. 1(c)). 
A general  distortion modifies both the 
radius and phase of the ${\tilde R}_j$'s; here we discuss the effect of each 
type of randomness separately.

In a phase--distorted chain at radius $\tilde{R}$, the site $\tilde{R}_{j}$ 
is given by
\beq
\tilde{R}_{j} = \tilde{R} \exp\{i ( j \theta + \delta_j)\}
\label{pdchain}
\eeq
where  $ \theta = \frac{2 \pi}{M} $ and $\delta_j$ is a random variable. 
Averaging over the distribution of $\delta_j$'s, we obtain
\bea
\langle\Sigma\rangle=\sum_{j=1}^{M}\langle \frac{R}{\tilde{R}_{j}}\rangle=
\nonumber \\
=\frac{R}{\tilde{R}}\langle \exp\{-i\delta\}\rangle
\sum_{j=1}^{M}\exp\{-i(j\theta)\}=0\; .
\label{avSig}
\eea
The lowest order contribution to $C_2$ (Eq. (\ref{c2def})) is then 
$\sim\langle|\Sigma |^2\rangle$, where 
\beq
\langle|\Sigma |^2 \rangle=\frac{R^2}{\tilde{R}^2}\sum_{j=1}^{M/2}
(2-2\langle \exp\{i\delta\}\rangle)\; .
\eeq
For a Gaussian distribution $P(\delta)=(1/\sqrt{2\pi}\sigma)
e^{(\delta^2/2\sigma^2)}$, we get
\beq
\langle|\Sigma |^2 \rangle=\frac{R^2}{\tilde{R}^2}M(1-\exp\{-\sigma^2/2\})
\approx\frac{R^2}{\tilde{R}^2}\frac{M\sigma^2}{2}\; ,
\label{avSigsq}
\eeq
where the last approximation holds for $\sigma\ll 1$. Substituting in Eqs.
 (\ref{c2def}) and (\ref{dosalph}), this yields
\beq
\alpha\approx\frac{L}{2\pi v}\frac{R^2}{\tilde{R}^2}\frac{M\sigma^2}{2}\; .
\label{alphrp}
\eeq

We next consider a radius--distortion of  Eq. (\ref{pchain}) of the form
\beq
\tilde{R}_{j} = (\tilde{R}+r_j) \exp\{i ( j \theta )\}\; ,
\label{rdchain}
\eeq
where $r_j$ is a random variable, subject to a distribution of width 
$\sigma_r\ll\tilde{R} $. Again, the first order in $\Sigma$ vanishes
upon averaging:
\beq
\langle\Sigma\rangle=\frac{R}{\tilde{R}^2}\langle r\rangle
\sum_{j=1}^{M}\exp\{-i(j\theta)\}=0\; .
\eeq
For a symmetric distribution of $r_j$'s, $\langle r\rangle=0$ and we get
\beq
\langle|\Sigma |^2 \rangle=\frac{R^2}{\tilde{R}^4}\sum_{j=1}^{M}
\langle r^2\rangle =\frac{R^2}{\tilde{R}^4}M\sigma_r^2\; .
\eeq
We then get an expression for $\alpha$ which is quite similar to Eq. 
(\ref{alphrp}):
\beq
\alpha\approx\frac{L}{2\pi v}\frac{R^2}{\tilde{R}^4}M\sigma_r^2\; .
\label{alphrr}
\eeq


\acknowledgements
We thank E. Akkermans, R. Ashoori, Y. Avron, J. Feinberg, Y. Meir and S.
Sondhi for fruitful discussions. 
This work was partly supported by grant no. 96--00294 
from the United States--Israel Binational Science Foundation (BSF), Jerusalem,
Israel, and the Technion -- Haifa University Collaborative Research Foundation.
M.M. also acknowledges support from
 the Fund for Promotion of Research at Technion, and the Israeli 
Academy of Sciences.

\appendix
\section{}
\label{app:A}
To evaluate $<\Psi(\xi)|\Psi(\xi)>$ (Eq. (\ref{plasmaint})), 
one needs to solve integrals of the form
\beq
I=\int d^2w f(w,w^*)|w|^{2mN}e^{-{1\over 2}|w|^2}e^{{1\over 2}wR_j^*}
e^{{1\over 2}w^*R_j}\; ,
\label{Iint}
\eeq 
where $w$ is the coordinate of the $j$'th WS electron (the index $j$ being 
omitted), and $f(w,w^*)$ can be expressed as a power series in $R/w,R/w^*$:
\beq
f(w,w^*)=\sum_{n,k=0}^\infty a_{nk}\left({R\over w}\right)^n
\left({R\over w^*}\right)^k\; .
\label{fww}
\eeq 
Eq. (\ref{Iint}) is then recast as
\bea
I=\sum_{n,k=0}^\infty a_{nk}R^{n+k}I_{mN-n,mN-k}\; ,\nonumber\\
I_{p,l}\equiv \int d^2w w^pw^{*l}e^{-{1\over 2}|w|^2}
e^{{1\over 2}wR_j^*}e^{{1\over 2}w^*R_j}\; .
\label{Ipl}
\eea
Below we show that provided $(mN-n),(mN-k)\gg 1$, 
\bea
I_{mN-n,mN-k}\approx  I_{mN,mN}{\tilde R}_j^{-n} ({\tilde R}_j^*)^{-k}\; ,
\nonumber\\ 
{{\tilde R}_j\over R_j}={{\tilde R}_j^*\over R_j^*}=e^{\alpha(|R_j|)}
\quad{\rm where}\quad \alpha(|R_j|)\approx {R^2\over |R_j|^2}\; ;
\label{Inkapp}
\eea
note that $\exp\{\alpha(|R_j|)\}\rightarrow 1$ for $(R/R_j)\ll 1$.
In addition, we note that since the integration over $w$ is dominated by
$w\sim R_j$, and $R_j>R$, the series expansion Eq. (\ref{fww}) can be cut at 
some $n_c,k_c$ such that $|R/R_j|^{n_c},|R/R_j|^{k_c}\ll 1$, yet 
 $(mN-n_c),(mN-k_c)\gg 1$. Consequently, we obtain 
\beq
I\approx f({\tilde R}_j,{\tilde R}_j^*)I_{mN,mN}\; ,
\label{Iapp}
\eeq
which implies that the $\{w_j\}$--integrations over the terms $x_{n}$ 
(Eq.  (\ref{sympol})) yield Eq. (\ref{Sympol}). 

We now derive the approximation  Eq. (\ref{Inkapp}) -- the central result of
this Appendix. The integrals $I_{p,l}$ (Eq. (\ref{Ipl})) can be expressed as
\bea
I_{p,l}=2^{p+l}{\partial^p\over \partial R_j^{*p}}
{\partial^l\over \partial R_j^{l}}\int d^2w e^{-{1\over 2}|w|^2}
e^{{1\over 2}wR_j^*}e^{{1\over 2}w^*R_j}= \nonumber\\
=2^{p+l}{\partial^p\over \partial R_j^{*p}}{\partial^l\over \partial R_j^{l}}
[4\pi e^{{1\over 2}|R_j|^2}]\; .
\label{Iplder}
\eea
A straight--forward application of the derivatives then yields
\bea
I_{p,l}=4\pi e^{{1\over 2}|R_j|^2}R_j^pR_j^{*l}S(p,l)\; ,\nonumber\\
{\rm where}\quad S(p,l)=\sum_{i=0}^{\min\{p,l\}}{p!l!\over i!(p-i)!(l-i)!}
\left({2\over |R_j|^2}\right)^i\; .
\label{ISpl}
\eea
We next assume that the above sum is dominated by $1\ll i\ll p,l$, so that
the factorials are well approximated by Stirling's formula:
\bea
S(p,l)\approx\sum_i s_i\; ,\nonumber\\
s_i=\left[{ple\over i}{2\over |R_j|^2}
\left(1-{i\over p}\right)\left(1-{i\over l}\right)\right]^i \times\\
\label{Spl}
\times \sqrt{{pl\over 2\pi i(p-i)(l-i)}}\; .\nonumber
\eea
The sum is then replaced by an integral
\bea
S(p,l)\approx{1\over \sqrt{2\pi}} \int dx&e^{\phi(x)}\; , \nonumber\\ 
\phi(x)=x\left\{\ln\left[\left({2ple\over |R_j|^2x}\right)
\left(1-{x\over p}\right)\left(1-{x\over l}\right)\right]\right\}
\label{Sint}\\
+O(\ln(x))\; ,\nonumber
\eea 
which can be solved in a saddle--point approximation. The saddle--point 
equation $\phi^\prime(x)=0$ implies 
\beq
{2ple\over |R_j|^2x}\left(1-{x\over p}\right)\left(1-{x\over l}\right)
=\exp\left\{1+{x\over p-x}+{x\over l-x}\right\}\; ,
\label{spt}
\eeq
and hence
\beq
S(p,l)\approx\sqrt{{1\over \phi^{\prime\prime}(x_s)}}
\exp\left\{x_s\left(1+{x_s\over p-x_s}+{x_s\over l-x_s}\right)\right\}
\label{Sspt}
\eeq
where $x_s$ is the solution of  Eq. (\ref{spt}).
For $(R/R_j)\ll 1$ (where $p,l\leq Nm=R^2/2$) 
\beq
x_s\approx {2pl\over |R_j|^2}\ll p,l\; ,\quad 
\phi^{\prime\prime}(x_s)\approx {1\over x_s}\; ,
\eeq
and we get
\beq
S(p,l)\approx\sqrt{x_s}e^{x_s}=\sqrt{{2pl\over |R_j|^2}}
\exp\left\{{2pl\over |R_j|^2}\right\}\; .
\label{Sfinal}
\eeq
Noting that $p,l\sim R^2/2$, Eq. (\ref{Sfinal}) implies  Eq. (\ref{Inkapp}).
{\it Q.E.D.}

\begin{figure}[htb]
\centerline{\psfig{figure=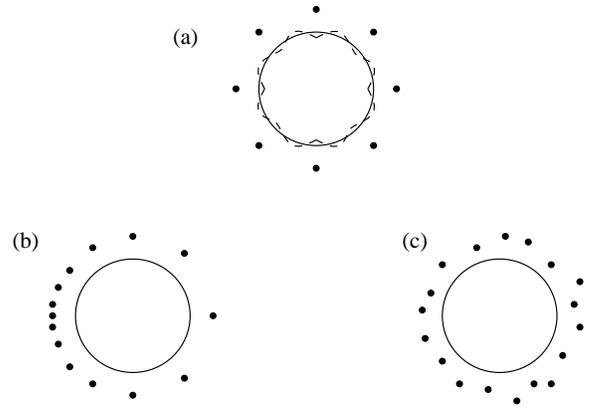,width=3in}}
\vskip0.5in
\caption[]
{
\label{fig:fig1}
A disc of QH liquid surrounded by a WS (the dots denote the sites at which the
WS electrons are localized). Three different types of site configurations
are sketched: (a) an ordered, commensurate chain (the effective boundary of
the liquid in the presence of the WS electrons is marked by a dashed line); 
(b) an inhomogeneous configuration; (c) a random distribution. 
}
\end{figure}


\begin{references}
\bibitem{fn1}
Throughout the paper we assume that the WS has time independent configuration. 
However, note that even if we consider (transverse and longitudinal)
propagating phonon excitations of a perfectly ordered configuration,
the conditions \ref{conditions} are still satisfied.
\end{references}
\end{document}